\documentclass[letterpaper,twocolumn,10pt]{article}

\usepackage{epsfig}

\usepackage{graphicx, subfigure}
\usepackage{verbatim}

\usepackage{algorithm}
\usepackage{algorithmic}
\usepackage{array}

\usepackage{amssymb}
\usepackage{array}
\usepackage{mathtools}
\usepackage{mathptmx}
\usepackage{xspace}
\usepackage{color}
\usepackage{epstopdf}
\usepackage{url}

\usepackage{perpage}	 

\MakePerPage{footnote}	 

\usepackage[hang,splitrule]{footmisc}

\usepackage{fancyhdr}
\fancypagestyle{plain}{%
\fancyhead[L]{Technical Report}
\fancyhead[R]{\thepage}
}

\newcommand{\cut}[1]{}

\newcommand{\note}[1]{{\color{red}[#1]}}

\usepackage[font=normalsize,labelfont=bf]{caption}

\usepackage{enumitem}

\makeatletter
\def\ps@headings{%
\def\@oddhead{\mbox{}\scriptsize\rightmark \hfil}%
\def\@evenhead{\scriptsize\thepage \hfil \leftmark\mbox{}}%
\def\@oddfoot{}%
\def\@evenfoot{}}
\makeatother

\newenvironment{noitemize}{
 \begin{list}{{\textbullet}}{
  \setlength{\partopsep}{0pt}
  \setlength{\parskip}{0pt}
  \setlength{\parsep}{0pt}
  \setlength{\topsep}{0pt}
  \setlength{\itemsep}{0pt}
  \setlength{\itemindent}{0pt}
  \setlength{\leftmargin}{9pt}
 }
}{
 \end{list}
}

\begin{document}

\title{SDNFV: Flexible and Dynamic Software Defined Control \\ of an Application- and Flow-Aware Data Plane}


\author{
{Wei Zhang${^*}$ \; Guyue Liu${^*}$ \; Ali Mohammadkhan${^\dagger}$\; Jinho Hwang${^\ddagger}$\; K.K. Ramakrishnan${^\dagger}$}\; Timothy Wood${^*}$ \\
\textit{${^*}$The George Washington University \; ${^\dagger}$University of California Riverside \; ${^\ddagger}$IBM Research}
} 

\date{}

\maketitle

\begin{abstract}

Software Defined Networking (SDN) promises greater flexibility for directing packet flows, and Network Function Virtualization promises to enable dynamic management of software-based network functions. However, the current divide between an intelligent control plane and an overly simple, stateless data plane results in the inability to exploit the flexibility of a software based network. In this paper we propose SDNFV, a framework that expands the capabilities of network processing-and-forwarding elements to flexibly manage packet flows, while retaining both a high performance data plane and an easily managed control plane.

SDNFV proposes a hierarchical control framework where decisions are made across the SDN controller, a host-level manager, and individual VMs to best exploit state available at each level. This increases the network's flexibility compared to existing SDNs where controllers often make decisions solely based on the first packet header of a flow. SDNFV intelligently places network services across hosts and connects them in sequential and parallel chains, giving both the SDN controller and individual network functions the ability to enhance and update flow rules to adapt to changing conditions. Our prototype demonstrates how to efficiently and flexibly reroute flows based on data plane state such as packet payloads and traffic characteristics.

\end{abstract}

\section{Introduction}
\label{section:introduction}

Software Defined Networking (SDN) is changing how networks are managed both in wide area networks and within data centers by providing a logically centralized control plane capable of applying carefully crafted rules that target individual packet flows~\cite{Feamster:2013:RS:2559899.2560327}.  At the same time, Network Function Virtualization (NFV) has emerged as a technique to run high performance network services as software running in virtual machines (VMs) on commodity servers~\cite{Hwang:2014:NHP:2616448.2616490,CLICKOS:NSDI:2014}, replacing custom-built middlebox systems commonly prevalent in networks. These technologies promise to increase the flexibility and agility of networks at both the control and data plane levels, allowing packets to be dynamically rerouted and new network services to be added on demand.

Unfortunately, the limited interactions between existing SDN controllers
and NFV platforms
restrict how far we can go in achieving this vision~\cite{Stratos:2012,SurveyMiddlebox:2012}.
The control plane, embodied by the SDN controller, has complete authority, but lacks insight into the information only observable within the network processing elements.
Information such as specific packet arrival sequences,
aggregate flow information, relationship of the state across flows, and application-level packet data is usually not exploitable by the SDN controller.
This
overly strict separation of the data and control planes limits both the knowledge on which control decisions can be made and the frequency with which the
handling of flows can be manipulated~\cite{Anwer:2013:SCP:2491185.2491223,Qazi:ONS:2013}.

While SDNs aim to provide convenient control over a minimalist data plane, the reality is that the functionality residing in a network's data plane is increasingly dynamic and complex. Today's networks are not built purely from switches and routers. They are also composed of proxies, caches, deep packet inspection systems, firewalls, policy engines and other middleboxes that act on individual packets to manipulate their routing, content, and QoS. The desire to efficiently
and flexibly
perform processing on packet data creates a tension that is not likely be resolved by current SDN approaches, 
since the match/action rule sets given to switches are not expressive enough for supporting complex functionality.
It is well-recognized that relying on the controller to frequently perform per-packet processing is too inefficient.
Moreover, current SDN designs 
do not help with dynamic management and modification of the rules for long-running flows (e.g., a video stream that should have its quality adapted over time based on available capacity).

We propose Software Defined Network Function Virtualization (SDNFV), a framework that expands the capabilities of network-processing elements to help manage and guide flows based on potentially more complex data plane information, while still retaining high performance and an easily managed control plane.
SDNFV expands on the existing protocols for a centralized
control plane, as embodied in proposed SDN approaches~\cite{NFV:Term:Web,Feamster:2013:RS:2559899.2560327}
and builds upon recent efforts in high performance NFV platforms supporting VM-based network services~\cite{Hwang:2014:NHP:2616448.2616490, CLICKOS:NSDI:2014, han2015softnic, garzarella2015virtual}.


In our SDNFV architecture, we use SDN and NFV controllers coordinated by an SDNFV Application.
The SDNFV Application defines complex network services as a graph of processing functions, and has a placement engine to decide how functions are instantiated in VMs across the network.
Services residing in our NF VMs are easy to develop, user-space applications that can maintain complex state about observed packet characteristics. The NF VMs are controlled by an NF Manager on each host.  SDNFV's focus is on how control should be split across this hierarchy--exploiting the application-awareness of individual NFs, the host-level resource management capabilities of the NF Manager, and the global view provided by the SDNFV Application.



We provide a cross-layer control protocol so that the network function VMs are able to update a flow's rules dynamically, subject to constraints imposed by the SDNFV Application. This allows SDNFV to base flow management decisions on characteristics that cannot be determined at flow startup or on traffic characteristics across multiple flows.
Finally, SDNFV can efficiently bring application-level awareness to the network, allowing layer 7 packet content to determine routing in ways not possible with existing SDN deployments.

The contributions of SDNFV are as follows:
\begin{itemize}
\item An architecture for integrating SDN and NFV to support complex service topologies composed from VMs running on one or more hosts, that can be dynamically instantiated and modified in response to network conditions.
\item An ILP optimization formulation for the NF placement problem that seeks to place VMs to minimize the maximum utilization of the network's links and NFV hosts, and determines the initial routes for flows.
\item Cross-layer flow management protocols that exploit state at the SDN controller, NF Manager, and VMs to allow routing based on information such as the current server load, characteristics of packet sequences, cross-flow aggregate information, and application-level packet data.
\item Optimizations on data plane processing such as automated load-balancing across VMs, flow entry caching for long service chains, and detection of services that can process a packet flow in parallel to reduce end-to-end latency.
\end{itemize}

We have implemented the SDNFV platform using a DPDK-based NFV platform for fast data plane processing, and the POX SDN controller.
Our experiments demonstrate several key use cases that illustrate the benefits of flexible flow management, such as dynamic treatment of video flows, denial of service mitigation, and `ant' and `elephant' traffic management.
We achieve two orders of magnitude higher throughput, and are able to dynamically respond to changing network policies far faster than existing approaches.

The paper is structured as follows: Section 2 provides background on SDN and NFV, and provides the use cases which motivate our work.  Section 3 describes how SDNFV's design provides hierarchical state management, service chain abstractions, and a placement algorithm.  Section 4 details SDNFV's implementation and systems-level optimizations.  We evaluate the platform's efficiency and effectiveness for a variety of use cases in Section 5 and present related work in Section 6.

\vspace{5mm}

\section{Background and Motivation}



\subsection{SDN and NFV}

SDN separates the network's control plane from the data plane, with a goal to enable flexible network and flow management by a logically centralized SDN controller. SDN applications perform traffic engineering, QoS, monitoring, and routing, and provide inputs to the SDN controller (through its North-bound interface). The SDN controller then signals network elements to configure flow tables via standardized protocols such as OpenFlow (via its South-bound interface) based on the first packet or a priori defined rules.
The data plane refers to these flow tables and sends packets according to the specified actions for matched rules~\cite{Ian:SDNOverview:2014,COMMAGAZINE:SDN:2013}.

NFV seeks to move network functions into software, rather than expensive, vendor-specific combinations of hardware and software. By using commercial off-the-shelf (COTS) servers and virtualization technologies, NFV provides network and cloud operators greater flexibility for using network-resident functions at lower cost (due to the use of standardized high-volume servers). SDN and NFV are complementary, having distinct roles~\cite{ETSI:NFVOverview:2014,ETSI:NFVUSECASES:2014}, with a shared goal of providing a more flexible and dynamic network.

The desire to add complex processing into the network is not new: two decades ago Active Networking sought to enable highly configurable functionality inside networks by allowing packets to be enhanced with code defining functionality that should be applied at selected nodes in the network~\cite{Tennenhouse:2007:TAN:1290168.1290180}.
However, networks have moved towards technologies offering simplicity and efficiency rather than flexibility, e.g., the universal focus on IP. Despite this, the need to have data plane functionality for security and QoS, whether at routers, switches, or middleboxes, has been well-recognized. Interdependence between packets, and having to be aware of sequences of packets is essential for middleboxes such as firewalls and proxies \cite{Paxon:2014}.

With SDNFV, we recognize that the data plane must do more than just forward packets, and seek to take advantage of SDNs and NFV to strike the right balance of {\em efficiency, flexibility, and ease of control}.
High speed NICs and multi-core servers let software-based network functions achieve performance far beyond what was previously possible, and the ability  to rapidly provision virtual functions on demand eliminates Active Networking's need for packets to include code within them. Finally, we rely on the centralized control of SDNs to provide convenient, network-wide management, while still permitting local control by NFs when forwarding depends on packet state. SDNFV provides the ability to dynamically
instantiate well-defined functions in the network to handle packet flows,
while reducing the burden on the SDN controller.

\begin{figure}[t]
\begin{center}
\includegraphics[width=3in, trim=0.01cm 0.01cm 0.01cm 0.01cm, clip=true]{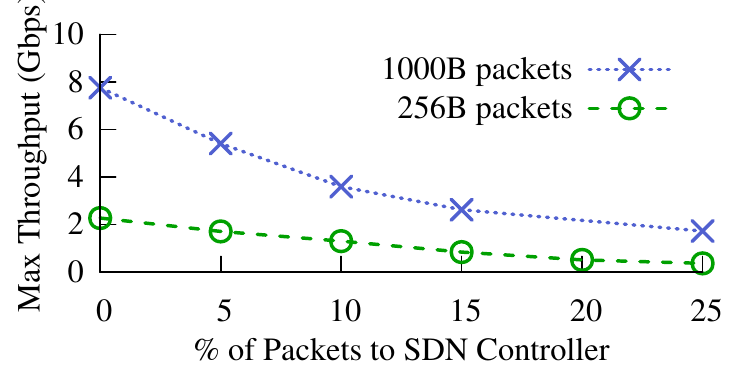}
\caption{Software switch performance quickly drops when the SDN controller becomes the bottleneck.}
\label{fig:background:sdncost}
\end{center}
\end{figure}


To evaluate the impact of coordination with the SDN controller, we measure the throughput achieved on the Open vSwitch (OVS) platform when a configurable percentage of traffic must be sent to the controller with packet size 256 bytes and 1000 bytes. OVS includes a software switch with a flow table; if there is a flow table miss, then a request is sent to the SDN controller.  Once a packet is matched it is simply sent back out the NIC.
As shown in Figure~\ref{fig:background:sdncost}, the maximum throughput that can be achieved quickly drops when the proportion of packets that must contact the controller increases. 
Here we use the POX controller, which is a single threaded python application, so it is the main performance bottleneck; however, we expect a similar trend even with higher performance SDN Controllers. 

Clearly, sending a significant portion of the network's traffic to the SDN controller will greatly reduce overall performance. To prevent this, many existing networks rely on proactively installed rules that match many flows, eliminating the need to frequently contact the controller.  While this lowers controller overheads, it also drastically reduces the flexibility the network can provide---decisions about packets must be made up front and cannot take advantage of the precise flow-level control SDNs can offer.
We believe the solution is not to conservatively pre-provision flow rules, nor is it to just make the controller faster~\cite{Dixit:2013:TED:2491185.2491193}. Instead we seek to increase the intelligence and independence of the data plane devices so they can directly manage specific flows, while following the guidelines provided by the control plane.

\subsection{Use Cases}
Here we describe two network service use cases that help motivate our design for SDNFV.

\vspace{1mm}
\noindent {\bf Anomaly Detection:} Networks are constantly under attack, so security systems must be highly {\em dynamic and efficient}--a difficult to achieve combination in today's world of hardware-based high performance middleboxes and slow software services.  SDNFV offers the opportunity to make network security services more flexible and reactive, while retaining high efficiency and low cost.  The system for security we envision combines an intrusion detection service (IDS) with distributed denial of service (DDoS) detection and mitigation.  Packets arrive at a {\em Firewall} NF and go through a {\em Sampler} NF, which takes a subset (either random or by shallow header inspection) of incoming traffic for further analysis.  The sampled traffic must then be processed by both an {\em IDS}, which looks for malicious signatures such as SQL exploits in HTTP packets, and a {\em DDoS Detector}, which coordinates with other DDoS Detector NF instances to find anomalous, possibly malicious, surges in traffic.  If either component finds suspicious traffic, all subsequent packets in the flow must be directed to a {\em Scrubber}, which performs a more detailed inspection of the packets to determine if they truly pose a threat to the network.

\vspace{1mm}
\noindent {\bf Video Optimization:} Video traffic consumes a significant part of the world's internet bandwidth.  Ideally, video flows should be automatically managed to try to provide both high quality of service and efficient, fair use of network bandwidth.  As part of this capability in the network, we consider a video transcoding service that adjusts video quality based on available resources.  Packets begin at a {\em Firewall}, then pass through a {\em Video Flow Detector} that analyzes HTTP headers of packets to detect the content type being transmitted in each flow.  Flows with video content are then sent to a {\em Policy Engine}, which considers factors such as available network bandwidth, time of day and financial agreements with consumers and/or video providers to decide whether this video's quality level should be adjusted. Flows that must be modified are passed to a {\em Quality Detector} that decides if the video can still retain the desired quality after transcoding.
The flow is passed to a {\em Transcoder} NF, which may use GPU processing to efficiently convert the video to a lower bit-rate.  The video flow passes through a {\em Cache} so that subsequent requests can be served locally, and finally exits through a traffic {\em Shaper}, which may limit the flow's rate to meet the desired network bandwidth level if necessary.

\begin{figure*}[t]
\begin{center}
\includegraphics[scale=0.55, trim=0.01cm 0.01cm 0.01cm 0.01cm, clip=true]{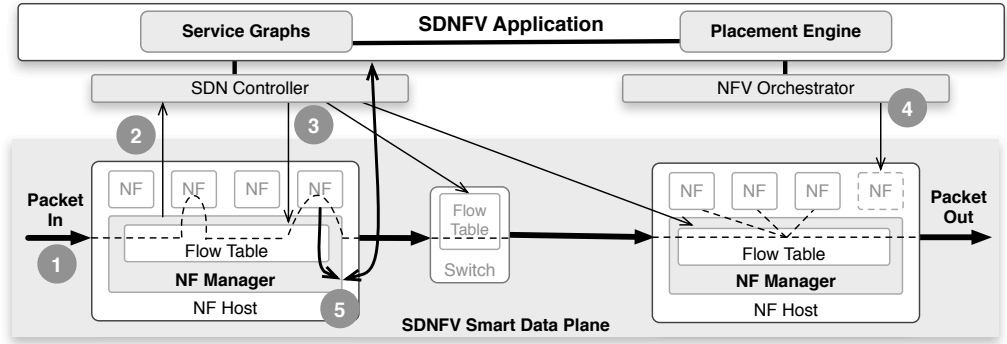}
\caption{The SDNFV Application's Service Graphs and Placement Engine guide the SDN Controller and NFV Orchestrator when configuring flow rules (steps 1-3) and instantiating new NFs (step 4). Based on packet data, an NF may request the NF Manager to adjust its flow rules, which requires communication with the SDNFV Application (step 5).}
\label{fig:sdnfv_flow}
\end{center}
\end{figure*}

\section{SDNFV Design}
\label{sec:design}
The two use cases in the previous section illustrate the complexity involved with typical network services--they are composed of multiple different components, and often data-plane information must affect how a flow traverses the NFs. This motivates the following design criteria for SDNFV:

\vspace{1mm}
\begin{noitemize}
\item Flexibly support complex network function service chains spanning multiple hosts.
\item Permit network functions to dynamically route flows, and enable flexible coordination between network functions and the SDNFV controller.
\item Support line rate packet processing with minimal latency, and provide scalability in multi-core deployments.
\end{noitemize}
\vspace{1mm}

We divide the control and state maintenance in the system into a three-layer hierarchy (\S \ref{sec:arch:state}).
This architecture enables the first two goals by representing the overall network service as a graph of network functions (\S \ref{sec:arch:servicegraphs}) that are then deployed inside a set of VMs spread across one or more hosts by a placement engine (\S \ref{sec:placement}). Our top level SDNFV Application coordinates with each host's NF Manager to determine the rules for a new flow, but allows VMs to update local flow table state as necessary (\S \ref{sec:decisions}). Finally, SDNFV is built on a zero-copy, polling-based packet processing framework that minimizes overheads when moving packets between network functions, and allows for maximal scalability by eliminating all critical sections (\S \ref{sec:implement}).
This design provides convenient high-level control in the SDNFV Application, but still leaves both the NF Manager and NFs some freedom in how individual packets are routed.  The key focus of our work is on providing the right abstractions and interfaces to allow this to happen flexibly and efficiently.

\subsection{State and Control Hierarchy}
\label{sec:arch:state}
SDNFV's 
goal is to limit state visibility to the tiers that can most efficiently gather it, while providing communication mechanisms to distribute information when needed.  One can classify the state being retained in a network's middleboxes based on the broad approach outlined in Split-Merge~\cite{Rajagopalan:2013:SSS:2482626.2482649}:

There is state that is {\em internal} to the NF and its host. We further categorize this as either NF-specific or host-specific.  NF-specific internal state includes transient state related to application logic or cached data, i.e., state which is necessary for the NF to run, but which does not need to be persisted or shared with other NFs or an SDN controller to help determine routing. The host-specific state covers metrics such as the occupancy of NF queues, the resource utilization of the host platform, and the list of NFs running on that host.


Next there is {\em external} state, which is related to the protocol and state that influences how packets are treated in the network. Some of this state has to be consistent across NF instances in the network, or even across different NFs that are involved in handling flows in the network~\cite{Rajagopalan:2013:SSS:2482626.2482649}. Split-Merge classifies data as ``Coherent" external state if it must be kept consistent across NFs, and ``Partitioned'' state, which is maintained independently for each NF.

In SDNFV, we do not explicitly deal with coordinating external state across NF replicas, instead our focus is on how the state managed at each layer guides flow routing decisions made across both the Control Plane and Data Plane in complex service chains. The hierarchical control framework we propose, illustrated in Figure~\ref{fig:sdnfv_flow}, bases its decision on both internal and external state maintained at each layer:

The {\bf SDNFV Application} has purview over the entire network, making it an ideal place to gather and maintain coarse grained, external state about flows, links, hosts, and NFs within the hosts.  It also holds the high-level policies that define complex services and their placement needs.

\vspace{1mm}
{\bf NF Managers} maintain and make decisions based on host-specific internal state about the resources available and the NFs running on the host. This allows NF Managers to track load levels of NFs for load balancing and respond to failure or overload. They also hold the flow table used on that host, which is initially provided by the SDN Controller and potentially updated
thereafter by the NFs.

\vspace{1mm}
The {\bf SDN Controller} and {\bf NFV Orchestrator} provide interfaces between the SDNFV Application and the NF Manager in order to setup flow table state and initialize NF VMs.

\vspace{1mm}
{\bf NFs} store application-specific internal and external state relevant for the flows they are tracking.  This often includes information gathered from payloads within each flow, for example an NF might parse an HTTP request in one packet and store information about the requested content to assist with processing subsequent packets (possibly affecting the behavior of other NFs). When an NF updates external state that should impact other NFs in the chain, we provide communication mechanisms for it to do so through the NF Manager.

Unlike other proposals, such as OpenNF~\cite{opennf}, where all of the non-NF specific state (both internal and external) is maintained and updated at the centralized SDN controller, we envisage a more distributed, hierarchical architecture for state maintenance. This relieves, potentially significant, overhead on the SDN controller. Further, it allows state maintenance at a finer granularity at the lower levels of the hierarchy, thereby obviating the need to frequently update the SDN controller and the associated communication overheads that often force protocol designers to settle for more coarse-grained state maintenance in a centralized approach.


\subsection{Network Service Abstractions}
\label{sec:arch:servicegraphs}

\begin{figure}[tb]
\begin{center}
\includegraphics[width=3.45in, trim=0.01cm 0.01cm 0.01cm 0.01cm, clip=true]{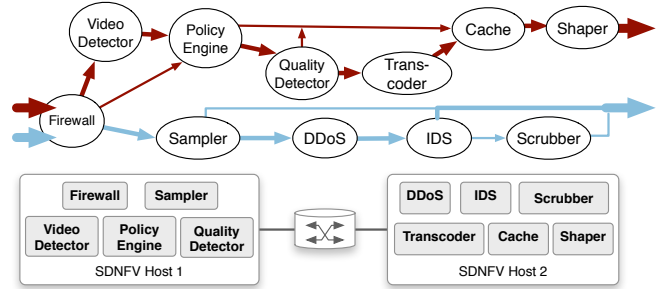}
\caption{Multiple graphs can be implemented onto the same hosts. The thick edges represent the default paths, which allow NFs to forward packets without knowledge of what services follow them. }
\label{fig:service-graph-map}
\end{center}
\end{figure}

\vspace{1mm}
\noindent {\bf Service Graphs} are used by SDNFV to simplify the description and management of applications like those described in the previous section.  A network application's processing requirements are represented by a graph with vertices for individual network functions and edges representing the logical links between them.
The graph defines the set of operations that can be applied to each type of flow as shown in Figure~\ref{fig:service-graph-map}.  This example shows how the two applications described previously can be represented as an abstract topology and how they can be implemented across multiple machines.

While network services are often thought of as ``service chains'' (i.e., linear sequences), we choose to represent each service graph as a DAG with a source and a sink.
This allows for more complex behavior than a linear chain, which is necessary if an NF may select a different next step in the graph depending on data inside the packet. The set of edges exiting a vertex define the possible next steps that an NF can specify when it completes processing a packet. This gives NFs fine-grained control, while still allowing the SDNFV Application to define the overarching functionality.

Network administrators define the service graph, and can specify one exiting edge from each vertex as the ``default'' path, shown by the thick edges in Figure~\ref{fig:service-graph-map}. An NF can decide {\em on a per-packet basis} whether to take the default path or a different edge. For example,
in the Anomaly Detection service graph, only sampled packets go through the DDoS and IDS NFs, while only those with suspicious content are sent to the scrubber.


\vspace{1mm}
\noindent {\bf Service IDs} provide SDNFV a layer of abstraction between a desired NF service (e.g., a Video Flow Detector), and the address of a specific VM that implements that service. This allows NFs to be replicated or moved throughout a network without having to modify the configuration of other functions that need to send packets to them (similar in philosophy to ~\cite{Arumai-ICN2015}). When an NF finishes with a packet, it can indicate that the packet should be sent to another Service ID; the NF manager on the host is then responsible for determining how the packet can be routed there, possibly after coordinating with the SDN controller.

\subsection{Instantiating Network Services}


\vspace{1mm}
\noindent {\bf From Service IDs to NFs:} Instantiating a service graph first requires deploying a set of NF VMs throughout the network using the placement algorithm described in \S \ref{sec:placement}. Since the service graph only dictates the abstract service types a packet must traverse, the NF Manager must determine which instance of the service to send a packet. If there are multiple VMs with the same Service ID residing on the same host, then the NF Manager can automatically load balance packets across them, either using knowledge of queue lengths to equalize load across VMs, or using hashing of the 5-tuple to ensure consistent processing of all packets in a flow.

\vspace{1mm}
\noindent {\bf Parallel packet processing} is another way the NF Manager can impact how a service graph is implemented.  Allowing several NFs to analyze a packet simultaneously can substantially decrease the latency for long chains.  However, since all parallel NFs are acting on one shared copy of the packet's data with no locks, different NFs must be read only, or must operate exclusively on different regions of the packet's data, e.g., one may rewrite the IP header while two other services simultaneously read the packet's body. Currently, SDNFV only permits read-only NFs parallel access to packets.

When a network function is started by the NFV Orchestrator, it must contact the host to setup its communication channels and advertise its service ID.
As part of this exchange it also specifies if it is a purely read-only service; the NF Manager then uses this information to determine if the service can be run in parallel with any adjacent NFs in the service graph.

For example, the DDoS and IDS services in Figure~\ref{fig:service-graph-map} only examine packets, without modifying their headers. All packets leaving DDoS go to the IDS, thus it is safe to permit both NFs to analyze the same packet at the same time. We discuss the actions that a network function can request for a completed packet in the next section and how concurrency issues are resolved in Section~\ref{sec:arch:optimizations}.

\vspace{1mm}
\noindent {\bf NF Manager Flow Tables:} The initial rules sent from the SDN Controller to each NF Manager are similar to the match/action rules defined by the OpenFlow protocol, but extended in two ways to support our service graph architecture: (1) In addition to the standard match fields, we include the Service ID the rule applies to, or a NIC port to represent rules for new packets. (2) For each match criteria we specify multiple actions and a flag indicating whether this is a list of parallel actions or a choice that will be made by the NF.  The list of parallel actions is used when there is a 
set of read-only NFs in the service graph.  In the non-parallel case, one action is listed for each edge in the service graph leaving the matched NF; this list represents the available next hops that the NF can request.
Our implementation easily makes these extensions to OpenFlow by repurposing the ``input port'' field in the match criteria to define service IDs, and by using OpenFlow's support for specifying multiple ``OUTPUT'' actions in a rule; we define the first action in the list to be the default action, and use a special action flag to indicate if the list is for parallel processing.


\subsection{Cross-Layer Flow Routing Protocols}
\label{sec:decisions}

\noindent {\bf Flow Table Setup:}
The SDNFV Application communicates with the SDN Controller via the controller's northbound interface to provide the flow rules that will forward packets between NFs as dictated by the service graph.  Rules can be determined on-demand when a new flow arrives, and then sent to all switches along the path in the service graph's implementation as shown in Figure~\ref{fig:sdnfv_flow} steps 2-3.  Alternatively, the SDNFV Application can use the SDN Controller to pre-populate rules---these could be designed to match a large category of flows (e.g., all flows originating from a set of video servers) in order to minimize flow table size.  Later, when a specific flow needs to be handled differently (e.g., a video flow to a mobile client that must be transcoded), new flow table entries will be created to define a new default path, {\em potentially without involving the SDN controller}.

\begin{figure}
\centering
\subfigure{
        \includegraphics[width=2.75in]{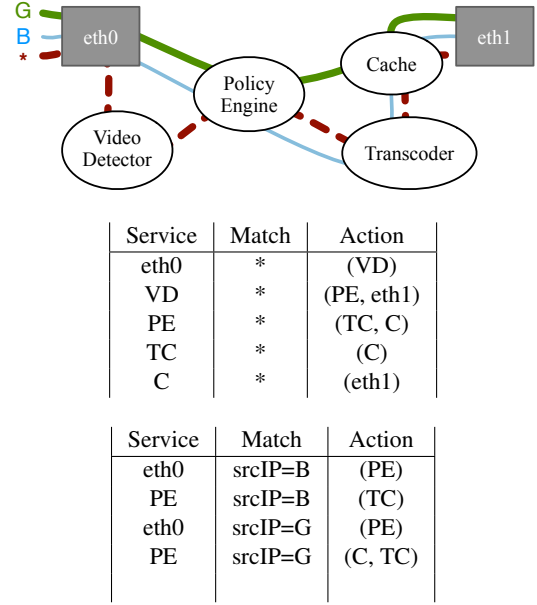}
}
\subfigure{
        {\small
        \begin{tabular}{|c|c|c|}
        Service & Match & Action \\ \hline
        eth0 & * & (VD) \\
        VD & * & (PE, eth1) \\
        PE & * & (TC, C) \\
        TC & * & (C) \\
        C & * & (eth1) \\
        \end{tabular}
        }
}
\subfigure{
        {\small
        \begin{tabular}{|c|c|c|}
        Service & Match & Action \\ \hline
        eth0 & srcIP=B & (PE) \\
        PE & srcIP=B & (TC) \\
        eth0 & srcIP=G & (PE) \\
        PE & srcIP=G & (C, TC) \\
          &  & \\
        \end{tabular}
        }
}
\caption{The service graph shows the default path for three different flows. The flow table at left shows the original rules determine by the SDNFV Application, while the table at right adds rules for specific flows.}
\label{fig:flow-examples}
\end{figure}

An example is shown in Figure~\ref{fig:flow-examples} for a simplified version of the Video Optimizer service. The initial flow table on the left holds rules to match all incoming flows ($*$). However,  two additional flows have been established and are given distinct rules (as listed in the table on the right) based on their specific needs. Note that we simplify the service graph illustration to only show the default (i.e., first) action, but an NF can still request a different action. For example, after some time the Policy Engine may redirect the Green flow to the transcoder instead of going directly to the cache.  Next we discuss how NFs perform these actions and how they impact the flow table rules.

\vspace{1mm}
\noindent {\bf NF Packet Actions:} When an NF finishes with a packet, it can invoke one of three actions: {\em Discard}, {\em Send to}, or {\em Default}. Discard drops the packet, while `Send to' causes the packet to be forwarded out a NIC port or to an NF with the specified Service ID--this is only permitted if the destination is one of the allowable next hops listed in the flow table derived from the service graph.   For most packets, the NF will invoke the Default action, which causes the packet's next step to be determined by the first flow table action associated with it.

Default actions are an important design point in SDNFV because they allow support for both tightly coupled and independent NFs. For example, a Firewall NF may have no knowledge of other NFs in the service graph. Rather than require it to be aware of the next hops, SDNFV lets it simply either drop packets based on its own rules or forward them using the default action specified by the SDNFV Application.  In contrast, an IDS NF might always be deployed as a pair with a Scrubber NF, thus it knows the service ID to send packets to, when it detects malicious content.

NFs do not directly forward packets, instead they set a meta-data flag  and return the packet to the NF Manager, which will then perform the requested action, if permitted. This allows the NF manager to translate Service IDs to addresses, or contact the SDN Controller to determine the next step if necessary.

\vspace{1mm}
\noindent {\bf Cross-Layer Control:} It is often desirable for an NF to be able to modify not only the next step taken by a flow, but the default action at previous or subsequent hops. To support this requires a communication protocol so that NFs can request changes from the NF Manager, and the NF Manager can in turn validate changes with the SDNFV Application.


SDNFV defines several messages that can be sent between NFs and the NF Manager. Each applies to flows matching some criteria, $F$, which can either be one single flow, or can use wild cards to match multiple flows:
\vspace{1mm}
\begin{noitemize}
\item {\em SkipMe($F$, $S$)} makes it so any NFs with default edges leading to $S$ will instead have a default edge leading to $S$'s default action, causing $S$ to be bypassed.
\item {\em RequestMe($F$, $S$)} makes all nodes that have an edge to $S$ set $S$ as their default action.
\item {\em ChangeDefault($F$, $S$, $T$)} updates the default flow table rule for service $S$ to be service $T$.
\item {\em Message($S$,$K$,$V$)} is used to send arbitrary {\em(key, value)} pairs from NF $S$ to the NF Manager.
\end{noitemize}
\vspace{1mm}

Compared to the packet actions described above, these messages have a longer lasting effect--they will adjust the flow table in the NF Manager (and potentially other hosts) so that all subsequent packets matching the specified flows will have new default rules. The $SkipMe$ and $RequestMe$ functions are provided since a loosely coupled NF may not know the complete service graph it is part of; e.g., the Video Detector NF in Figure~\ref{fig:service-graph-map} may want to adjust the rules for a flow once it detects video content, but it may not know what type of service precedes it.  Alternatively, a security appliance may need to make substantial changes to the service graph once it detects a coordinated attack; it can do so using the $ChangeDefault$ messages, albeit constrained to edges defined in the original service graph.

The $Message$ call is used to send arbitrary application data between layers. We assume the service ID, $S$, is used to determine the appropriate message handler component that can interpret the application-specific message.
This could potentially invoke logic in the SDNFV Application to completely rewire the service graph, or it could be used to simply update global state.

All of the above message types are first sent from the NF to the NF Manager, which can then pass them to the SDNFV application either for validation (if NFs are not fully trusted) or for messages that would affect other hosts.  Communication to the SDNFV Application could be done either directly or through the SDN Controller (as shown in Figure~\ref{fig:sdnfv_flow} step 5). The latter may entail enhancements to the OpenFlow protocol.

\vspace{1mm}
\noindent {\bf NF--SDN Coordination:} Given the ability of NFs and the NF Manager to manipulate flow routing and instantiate new replicas, SDNFV needs a mechanism to push information back to the SDNFV Application and the SDN Controller so they can be made aware of these changes when necessary. In our current implementation, we rely on our $Message$ command to move data from an NF to the NF Manager, and then to the SDNFV Application, which can manipulate the SDN Controller over its north-bound interface. However, we believe that in the future, NF--SDN communication protocols such as OpenFlow must evolve to better support information flow {\em from the NF to the SDN controller}, a direction that currently has only rudimentary support. This would allow NFs to provide generic statistics such as flow or drop rates, as well as application specific triggers such as detection of malicious flows or component failures.

\subsection{NF Placement}
\label{sec:placement}

The SDNFV Application involved in orchestration of the NFs in the network takes advantage of a Placement Engine that solves an optimization formulation to determine the number of instances of particular NFs and their locations for a given network topology and a set of flows. It concurrently determines the route for those flows.
The optimization accounts for a number of constraints, with the goal of minimizing resource utilization (thus maximizing the number of flows the system can support). The
formulation also enables us to develop heuristics that make the approach practically useful. Our heuristic solves the placement and routing of a set of flows through the required service chain of NFs, and then incrementally solves for additional flows based on the residual capacity, without impacting the already assigned flows. Unlike previous work~\cite{Sekar:2012:DIC:2228298.2228331}, our approach solves the joint problem of placement of the necessary instances of NFs \emph{and} the routing of flows through the NFs, based on the service graph specification.
We formulate this as a mixed integer linear program (MILP). Notation is in Table~\ref{table:ILP}.

{\em Minimize U subject to:}\\
\setlength{\jot}{0mm}
\begin{gather}
\forall{k \in Flows},\forall{i,m \in Switches},\forall{j \in Services},\forall{l \in O_k},\nonumber\\
\forall{l' \in O'_k}:   \nonumber\\
\sum\limits_{j} M_{ij} \leq C_{i} \label{eq:cores}\\
N_{k_{il}} = M_{ij} S_{k_{jl}} W_{k_{il}}\label{eq:mweq}\\
\sum\limits_{i} N_{k_{il}} = 1 \label{eq:OneInstance}\\
F_k = [ I_k N_k E_k]\label{eq:entryExit}\\
\sum\limits_{m} V_{k_{l'_{im}}} - \sum\limits_{m} V_{k_{l'_{mi}}} = F_{k_{il'}} - F_{k_{i(l'+1)}}\label{eq:shortestpath}\\
\sum\limits_{l'} \sum\limits_{i,m} V_{k_{l'_{im}}} D_{im} \leq T_k \label{eq:delayEnforce}\\
\sum\limits_{k} S_{k}N_{k}^T \leq P_{ji} \label{eq:maxCap}\\
\sum\limits_{k} \sum\limits_{l'} (V_{k_{l'_{im}}} B_k)/H_{im} \leq U\label{eq:linkUtil}\\
\sum\limits_{k} S_{k}N_{k}^T / P_{ji} \leq U \label{eq:APU}
\end{gather}
Equation~\eqref{eq:cores} reflects the constraint on the number of available processor cores on an NF Host. We assume services do not share CPU cores to avoid context switching overheads. Equation~\eqref{eq:mweq} shows the selected nodes for flow $k$
and Equation~\eqref{eq:OneInstance} ensures that one node is selected for each service for flow $k$. Equation~\eqref{eq:entryExit} adds the flow's entry and exit nodes to the selected path.
 Equation~\eqref{eq:shortestpath} selects the links to the next node in the service chain, with only the
source (+1) and destination (-1) nodes having a non-zero value. Equation~\eqref{eq:delayEnforce} ensures that the selected routes do not exceed the maximum tolerable delay for a flow. The maximum capacity of each core is enforced by Equation~\eqref{eq:maxCap}. Finally, Equations~\eqref{eq:linkUtil} and~\eqref{eq:APU} constrain the maximum utilization of links and cores.
The core utilization is measured as a function of the maximum number of flows supported by a CPU core implementing a particular service.

\begin{table}[tb]
        \begin{center}
                \footnotesize
                                \begin{tabular}{p{0.3cm}|p{7.8cm}}
                                        \textbf{Var.}  & \textbf{Definition } \\  \hline
                                        $U$ & Maximum utilization of links and switches\\
                                        $M_{ij}$ & Number of running instances of service $j$ on switch $i$\\
                                        $C_i$ & Number of available cores on switch $i$\\
                                        $N_{k_{il}}$ & Selected switch for order $l$ of the flow $k$'s service chain\\
                                        $S_{k_{ij}}$ & 1 if $i$ is the $j^{th}$ service in flow $k$'s service chain; 0 otherwise\\
                                        $W_k$ & Binary decision variable for satisfying constraint \eqref{eq:OneInstance}\\
                                        $I_k{_{i}}$ & 1 if $i$ is the entrance switch for flow $k$\\
                                        $E_k{_{i}}$ & 1 if $i$ is the exit switch for flow $k$\\
                                        $D_{ij}$ & Delay of the link between switches $i$ and $j$\\
                                        $T_K$ & Max. delay tolerated by flow $k$\\
                                        $P_{ij}$ & Max. number of supported flows, if switch $i$ runs service $j$\\
                                        $B_K$ & Bandwidth usage of flow $k$\\
                                        $H_{ij}$ & Capacity of the link between switch $i$ and $j$\\
                                        $O_k$ & A range from 1 to length of service chain for flow $k$\\
                                        $O'_k$ & A range from 1 to length of service chain for flow $k$ plus one \\

                                        $V_{k_{l_{im}}}$ & Is one, if the link between switches $i$ and $m$ is used, to reach to the $l^{th}$ service in service chain of flow $k$\\

                                \end{tabular}
                \caption{Definition of variables of MILP formulation}
                \label{table:ILP}
        \end{center}
\end{table}

We compared the placement based on the MILP solution with that of a greedy best effort heuristic that assigns services to the first available cores on network nodes in the shortest path
for the flow, and, if needed uses additional cores on neighboring nodes on the flow's path.
With our heuristic, we divide the problem into sub-problems, each having a small number of flows (e.g., 5) so as to compute the solution quickly. After solving a sub-problem,  a post processing step updates the available resources, removes unused switches and solves the next sub-problem for the next
small subset of flows. We continue this process until services are assigned  to all requesting flows at any point in time. We call this `Division Heuristic'.
We ran all the algorithms on a particular network topology (AS-16631 from Rocketfuel~\cite{neil2003quantifying} with 22 nodes and 64 edges). All flows have a similar service chain with a length of 5 ($J_1$-$J_5$) services.  All nodes are homogeneous and each have 2 CPU cores. Each core supports up to 10 flows for service $J_1$-$J_4$ and 4 flows for $J_5$.

\begin{figure}[t]
        \begin{center}
                \includegraphics[width=3in]{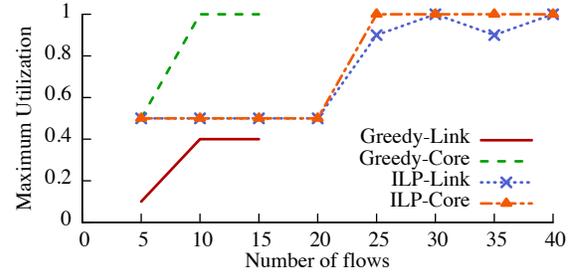}
                \caption{Maximum number of supported flows by different algorithms (Greedy, Optimal, and Division Heuristic) for different network capacities (1-100X of the original CPU and link capacity)}
                \label{fig:utilization}

        \end{center}
\end{figure}

The left sub-figure in Figure~\ref{fig:utilization} shows that the greedy heuristic is inefficient, with substantially fewer flows being accommodated. Solving the MILP optimally, on the other hand,  accommodates 3 times as many flows.
However solving the problem optimally for a large numbers of flows is not tractable. Using the division heuristic, we can fit about 85\% of the flows accommodated by the optimal solution, while taking less than a minute of computation on a laptop.
The right sub-figure shows that division heuristic is scalable. When we increase the capacity of links and the CPU, we can support a much larger number of flows in the network, and perform better than the greedy heuristic. Solving the formulation optimally takes too long and isn't scalable when the number of flows goes beyond a hundred.

\section{Implementation and Optimizations}
\label{sec:implement}
We have implemented our SDNFV architecture using the KVM/QEMU virtualization platform and the POX SDN controller. The enhancements we propose could be similarly applied to more recent controllers with updated OpenFlow protocols. Our implementation builds upon our NetVM high performance NFV platform~\cite{Hwang:2014:NHP:2616448.2616490}; the techniques could also be applied to platforms such as~\cite{CLICKOS:NSDI:2014, han2015softnic, garzarella2015virtual}. We extend NetVM with hierarchical flow management, SDN communication channels, and optimizations such as parallel processing.

\subsection{NF Manager}

The core component running on each host is the NF Manager.  This software runs as a user space application in the host OS, and manages the delivery of packets between network function VMs and maintains the flow table.

SDNFV relies on {\bf zero-copy communication channels} to exchange packets and other messages between virtual machines and the host. To support line rates of 10 Gbps or higher, synchronization primitives such as locks cannot be used since they can take tens of nanoseconds to acquire even in the uncontested case~\cite{IntelGettingStarted:2013,David:2007:CSO:1281700.1281703}. Therefore, we implement all communication in our system using asynchronous ring buffers that hold small messages, and large shared memory regions which can be referenced in these messages to provide zero-copy access to larger data structures when necessary.  Since each ring buffer has a single data producer thread and a single consumer thread, no locks are required to ensure consistency.

Zero-copy communication of network packets is achieved by having DPDK DMA packets into huge pages allocated within the host. These huge pages are then shared with the VMs, and lightweight ``packet descriptors'' are exchanged between domains using ring buffers to control who has access to a packet.  The NF Manager maintains a pair of ring buffers for each VM: one for sending packets to the VM and one for receiving packets from the VM.


When a packet arrives to an NF Host, it is received by an {\bf RX Thread} using DPDK's poll mode driver to prevent interrupt and system call overheads.  The thread queries the flow table to determine which VM(s) should receive the packet next and adds a small packet descriptor to that VM's RX ring buffer.

Alternatively, if this is the first packet for a flow, then the flow table miss causes the packet to be added to the RX ring buffer of the {\bf Flow Controller Thread}.  The Flow Controller will buffer the packet and send its header to the SDN controller over a secure channel.
We repurpose the OpenFlow protocol's OFPT\underline{~~}FLOW\underline{~~}MOD messages to define the forwarding actions between network functions. The SDN Controller will return a set of actions that implement the portion of the service graph contained on this host, as well as a forwarding rule so packets from the last VM in the host will be properly forwarded to the next host that implements a portion of the service graph (or a switch en route to that host). We consider each NF instance as a logical network port to which packets can be sent using standard open flow rules. When the SDN Controller wants a flow to be processed by a network function with service id $SID$, it sets the action for the flow to be ``output to port $SID$''.

When a network function finishes with the packet, it adds it to a TX ring buffer shared with the NF Manager. The NF Manager's {\bf TX Thread} reads the packet descriptor from the buffer and will do the action specified by the NF (e.g., discard it or send it to a specific Service ID), or perform another flow table lookup.  The TX Thread then passes the packet on to the appropriate NF, drops it, or sends it out the designated network port.


\subsection{NF Manager Optimizations}
\label{sec:arch:optimizations}

Our NF Manager includes several optimizations to minimize latency and maximize throughput.

\vspace{1mm}
\noindent {\bf Parallel Packet Processing:}
To support parallel access by multiple VMs, we extend the packet data structure used by DPDK to include a reference counter indicating how many VMs are concurrently accessing the packet.  The RX Thread increments the reference counter by the parallelization factor.
The packet descriptor is then copied into the RX buffers of all NFs in the action list, allowing them to access the shared memory in parallel.

If a packet is being processed by multiple VMs, it is possible that each will request a different action when they finish. The NF Manager's TX thread resolves conflicting action requests by either prioritizing actions (e.g., drop is most important, followed by transmit out, etc), or by having priorities associated with each VM (i.e., the action of the highest priority VM always takes precedence).

\vspace{1mm}
\noindent {\bf Caching flow table lookups:} Extracting the required meta data from a packet header and performing a flow table lookup can be costly if it is performed at each step in a long service chain.  To reduce this performance cost, we propose caching the flow table lookup result inside the packet descriptor that is transmitted between ring buffers.  This avoids the need for the NF Manager's TX thread to make hash table lookups, thus reducing latency.


\vspace{1mm}
\noindent {\bf Automatic Load Balancing:}
Particularly when running network functions that involve complex payload analysis, the amount of time to process a packet can vary on a per-packet basis. As a result, using round robin load balancing of packets to NFs can lead to unbalanced queue sizes, potentially leading to packet drops or variable latency.

To avoid this, the NF Manager can perform state-based load balancing based on the number of occupied slots in each VM's ring buffer.
Unfortunately, this approach cannot be used for NFs that involve temporal flow state, since this is typically kept on a per-thread basis to prevent concurrency issues. Therefore, the NF Manager can also perform load balancing based on the flow key, ensuring that all packets for a flow are sent through the same NF threads (unlike round robin or queue-size based balancing).

\subsection{Network Function VMs}
\label{sec:arch:nfv}

Each VM runs a single network function as a standard user space application linked to our SDNFV-User library. By default, a VM is assigned one core per socket (to prevent NUMA overheads), but additional cores can be added. Each core runs a thread implementing the desired network function, and has its own ring buffers shared with the RX and TX threads in the host on the same CPU socket.

The SDNFV-User library initializes the ring buffers and shared memory regions. It then advertises to the NF Manager what network service is offered by the VM. When a packet arrives in a thread's ring buffer, the library calls the custom function that implements the desired packet processing.  The library also provides a variety of helper functions for working with packet data, such as finding different headers or manipulating source/destination IP addresses.  This allows complex services to be written with minimal code.


\section{Evaluation}
\label{sec:eval}

\begin{figure}[t]
\begin{center}
\includegraphics[width=2.7in, trim=0.01cm 0.01cm 0.01cm 0.01cm, clip=true]{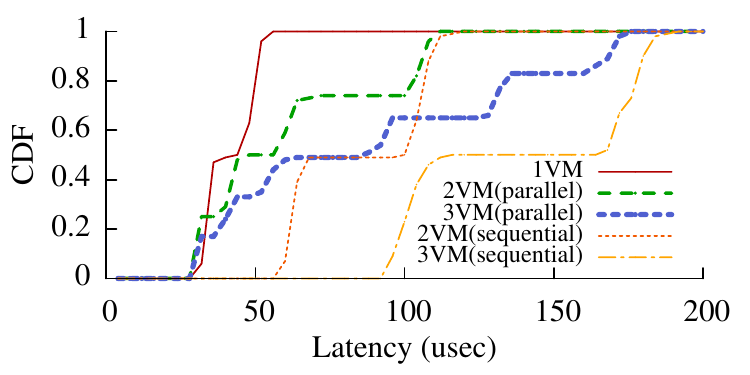}
\caption{Average roundtrip latency for VMs perform internsive computation on each packet}
\label{fig:expt:latency}
\end{center}
\end{figure}

We evaluate SDNFV to demonstrate how it can effectively provide an efficient and flexible data plane.


In our experimental setup, we use four servers with dual Xeon X5650 @ 2.67GHz CPUs (2x6 cores)---two for
the system under test and the other two acting as a traffic generator---each of which has an Intel
82599EB 10G Dual Port NIC (with one port used for our performance experiments) and 48GB memory,
each with 8GB for huge pages.
The host OS is Ubuntu 12.10 (kernel 3.5), and the guest OS is Ubuntu 12.10
(kernel 3.5). DPDK-1.4.1 and QEMU-1.5.0 are used.
We use PktGen-DPDK from WindRiver to generate traffic~\cite{PktGen:2013}.
The base core assignment, unless stated otherwise, uses 2 cores to receive, 4 cores to transmit/forward, and two cores per VM, all evenly split over two sockets.

\subsection{Sequential vs Parallel Processing}
\label{sec:eval:parallel}

SDNFV's goal is to provide a high performance data plane to build complex network functions. We measure both throughput and latency with varying number of VMs composed either sequentially or in parallel. We preconfigure the flow table rules to prevent SDN lookups affecting performance.

\vspace{1mm}
\noindent {\bf Latency}:  To quantify latency, we use a traffic generator to send 1000-byte packets at a low rate (100 Mbps) to the SDNFV platform. After being processed by one or more VMs, the packets are returned to the generator, which then measures the average roundtrip latency by checking a timestamp included in the packet.  Table~\ref{table:latency} shows the latency for different numbers of VMs running a no-op application that performs no processing on each packet. To show that SDNFV achieves comparably low latency, we also measure a simple DPDK forwarding application that doesn't involve any virtualization overheads. The results show that both parallel processing and inter-VM delivery introduce minimal additional latency and SDNFV is very efficient when VMs have negligible processing.

\begin{table}[b]
\begin{center}
\small
\begin{tabular}{c|c|c|c}
\textbf{\#VM \textbackslash Latency (us)  } & \textbf{Avg } & \textbf{Min } & \textbf{Max } \\ \hline
0VM (dpdk)                                & 26.66        & 23           & 29           \\
1VM                                      & 27.78        & 24           & 35           \\
2VM (parallel)                            & 28.02        & 24           & 37           \\
3VM (parallel)                            & 28.38        & 24           & 40           \\
2VM (sequential)                           & 28.86        & 24           & 43           \\
3VM (sequential)                           & 29.96        & 25           & 49           \\
\end{tabular}
\caption{Average roundtrip latency for VMs perform no processing on each packet}
\label{table:latency}
\end{center}
\end{table}

\begin{figure}[t]
\begin{center}
\includegraphics[width=2.7in, trim=0.01cm 0.01cm 0.01cm 0.01cm, clip=true]{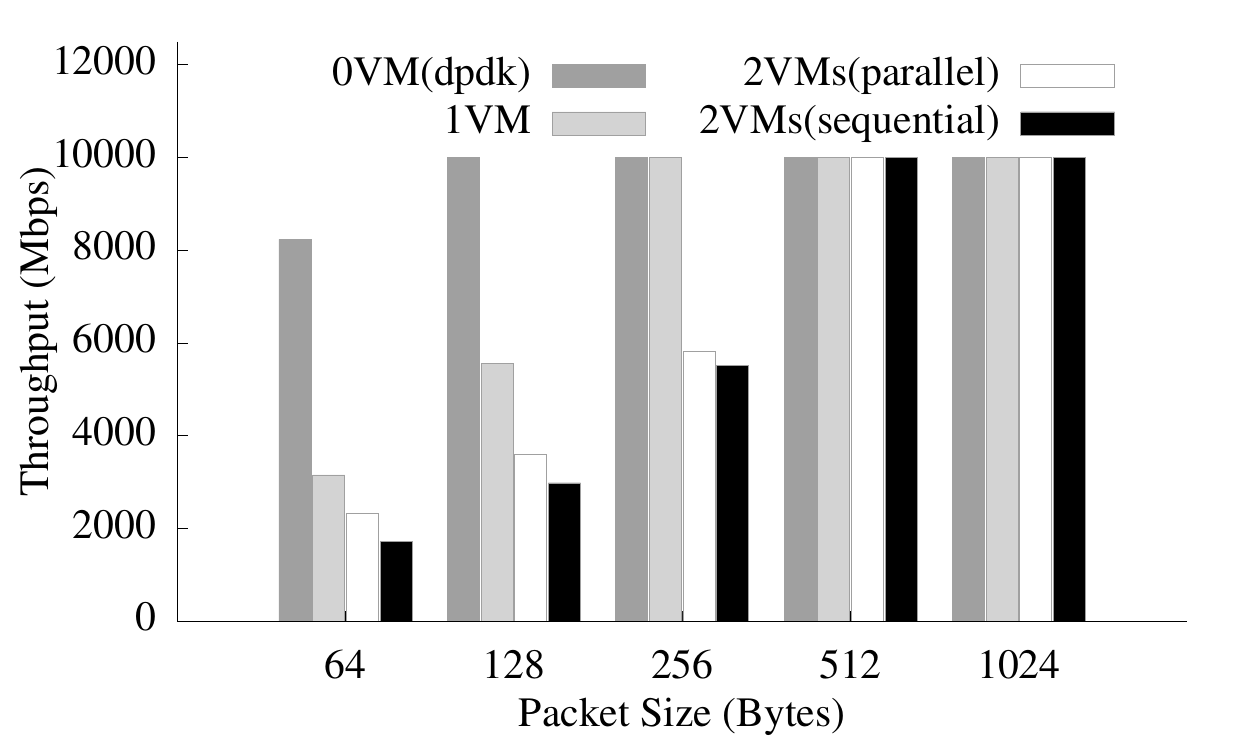}
\caption{Throughput achieved by SDNFV when adjusting the service chain length and level of parallelism with different packet size}
\label{fig:expt:throughput}
\end{center}
\end{figure}

\vspace{1mm}
To further show the efficiency of our parallel processing, we measure the latency when each VM performs an intensive computation on each packet. Figure~\ref{fig:expt:latency} shows the CDF of latency for different cases. All CDFs are based on the average across three runs. As expected, the results show parallelism can reduce the latency caused by long chains that include expensive VM processing.

\vspace{1mm}
\noindent {\bf Throughput}: To evaluate how pipelining and flow table lookups affect throughput, we measure the achieved throughput when adjusting the service chain length and level of parallelism. For this experiment we use a single
CPU socket. The NF Manager is allocated one RX thread, two TX threads (to prevent inter-VM or external transmissions from being the bottleneck), one flow controller thread, and one VM manager thread. Additionally, we use one thread per VM. All threads are pinned to a dedicated core, except for the VM manager and Flow controller which share a core on the second socket. On our platform, this allows us to run at most two VMs before running out of cores on the CPU socket; enabling the second CPU socket can double performance since the NIC splits the traffic evenly between the two.

Figure~\ref{fig:expt:throughput} shows the receive rate measured by the traffic generator when varying the packet size from 64 bytes to 1KB. We compare SDNFV with the simple DPDK forwarder which can achieve line-rate for most packet sizes, but does not use virtual machines for processing.  SDNFV can achieve close to 5Gbps for smaller packet sizes when using one socket and can achieve 10Gbps with larger packet sizes. More powerful systems should easily be able to support more VMs and achieve the full-line rate for all packet sizes using SDNFV.

\vspace{1mm}
\noindent {\bf Flow Lookups:} We next measure the overhead of flow management operations. We find that a Flow Table lookup takes an average of 30 nanoseconds, and the NF Manager can determine the VM with minimum queue sizes in 15 nanoseconds.  Performing an SDN lookup takes an average of 31 milliseconds, but this is deferred from the critical path to the Flow Controller thread, which makes those requests asynchronously.




\begin{figure}[t]
\begin{center}
\includegraphics[width=2.7in, trim=0.01cm 0.01cm 0.01cm 0.01cm, clip=true]{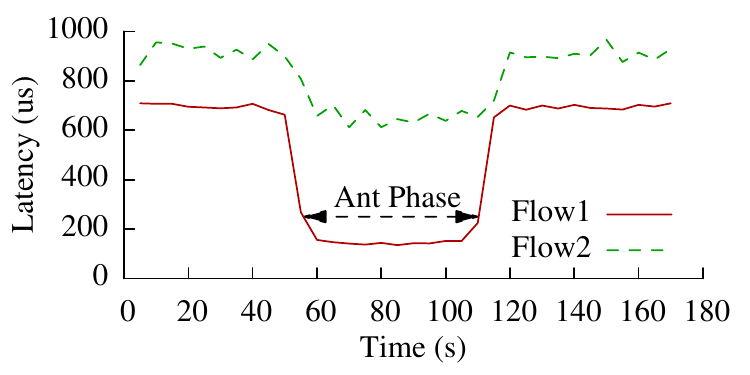}
\caption{SDNFV detects flow types and redirects ant flow to faster link}
\label{fig:expt:elephant}
\end{center}
\end{figure}

\subsection{Single and Cross Flow Management}
\label{sec:eval:ant}

\noindent {\bf Ant Flow Detector:} This  NF monitors long lived flows to determine their traffic behavior. Once detected, the service chain is modified so ant flows will see lower latency, representing a QoS management system differentiating between bulk transfers and latency-sensitive multi-media traffic.

Our Ant Detector NF classifies incoming flows by observing the size and rate of packets over a two second time interval. We use Pktgen-DPDK to emulate two flows: Flow 1 has small packet size (64 bytes) with varying traffic rate, Flow 2 has a large packet size (1024 bytes) with a constant rate. In Figure ~\ref{fig:expt:elephant},  both flows have high traffic rate from time 0s to 50s. They are both classified as elephant flows, so they see high latency. From 51 to 105s, Flow 1 reduces its rate. The NF reclassifies the flow because of this phase change and adjusts the default path to a higher speed link by sending a {\em ChangeDefault} message to the NF manager to update the default path.  This new path substantially lowers the ant flow's latency, but also improves Flow 2, since there is less network contention. After 105s, Flow 1 increases the rate back up. The NF reclassifies Flow 1 as an elephant, causing its latency to rise again.


\noindent {\bf DDoS Detector:} It is important to let the data plane not only characterize flows based on sequences of packets, but also manage traffic using aggregated information across multiple flows. This experiment shows that SDNFV can achieve multi-flow aware traffic management without expensive coordination with the SDN controller. We mimic a DDoS detection and mitigation system where initially a single VM monitors packets to detect potential attacks.  For simplicity, it detects an attack when the volume of traffic from an IP prefix exceeds a threshold, at which point the NF requests that a second DDoS Scrubber VM be created and all traffic be routed to it for cleaning. Our scrubber drops the malicious traffic.

\begin{figure}[t]
\begin{center}
\includegraphics[width=2.7in, trim=0.01cm 0.01cm 0.01cm 0.01cm, clip=true]{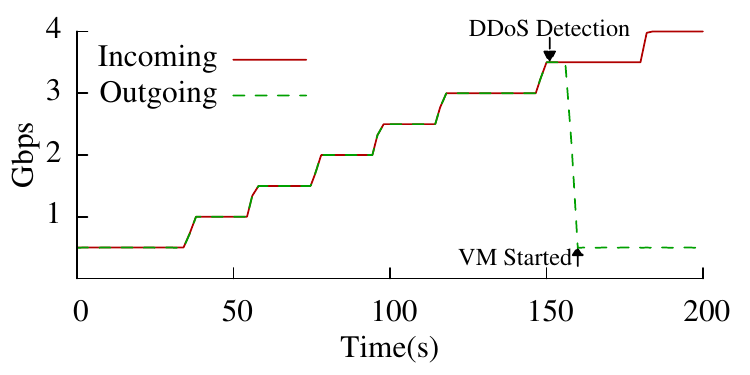}
\caption{SDNFV supports multi-flow aware flow traffic management, flexibly rerouting traffic to a DDoS Scrubber VM that drops the malicious traffic}
\label{fig:expt:ddosl}
\end{center}
\end{figure}

We use two traffic generators to generate normal and DDoS traffic separately. Figure~\ref{fig:expt:ddosl}  shows the outgoing and incoming traffic through the SDNFV over time. We send normal traffic at a constant rate (500Mbps), and start sending low rate DDoS traffic at 30s. Our DDoS Detector VM monitors incoming traffic patterns within a monitoring window and gathers data across all flows. We note the incoming traffic gradually rises until reaching our threshold (3.2Gbps), at which point it is detected as a dangerous level of suspicious traffic (this VM could incorporate complex anomaly detection algorithms). The NF uses the {\em Message} call to propagate this alarm through the NF Manager to the SDNFV Application. This causes a new DDoS Scrubber VM to be automatically started. The Scrubber VM sends the message {\em RequestMe} to the NF manager, which adjusts the default action so subsequent packets from the DDoS Detector VM are rerouted to the scrubber.
As expected, the outgoing traffic will return to normal levels as the scrubber drops malicious packets, even as the incoming traffic continues to rise.

This experiment shows how SDNFV can dynamically modify the service graph used for a flow by initializing new VMs and rerouting traffic. The time cost to boot up a new VM is about 7.75 seconds, but this cost can be further reduced by just starting a new process in a stand-by VM or by using fast VM restore techniques~\cite{Lagar-Cavilla:2009:SRV:1519065.1519067}.

\begin{figure}[t]
\begin{center}
\includegraphics[width=2.7in, trim=0.01cm 0.01cm 0.01cm 0.01cm, clip=true]{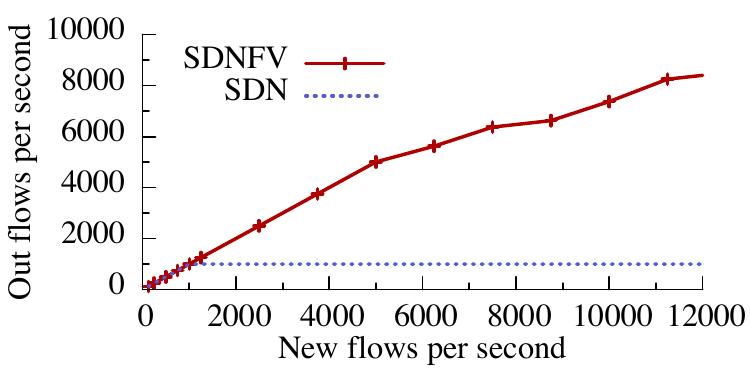}
\caption{SDNFV scales by distributing control decisions.}
\label{fig:expt:newflows}
\end{center}
\end{figure}

\subsection{Dynamic Video Flow Management}
Next we emulate a video service to evaluate SDNFV's performance overhead and dynamic flow control. Our video service is composed of three components: Video Detector, Policy Engine and Transcoder. The first two components make flow decisions dynamically based on the packet content and network state, and decide whether flows are routed to the transcoder which emulates down sampling by dropping packets. The goal of the system is to adjust the overall video traffic rate to meet a target bandwidth. We compare a centralized SDN approach to SDNFV, which achieves better efficiency and flexibility by distributing control to the NFs.

In current SDNs, the Video Detector and Policy Engine must be integrated into the SDN controller itself because only the controller has decision making power over flows. As a result, the first two packets of each flow from the web server to the client (i.e., the TCP connection ACK and the HTTP reply) must be sent to the SDN controller. It examines the payload in the second packet before setting up a flow table rule in the NF Manager to send video flows either directly out of the host or through the transcoder depending on the current policy.

For the SDNFV case, all three components are running in isolated VMs on one host managed by the NF manager. The header of the first packet in a flow is sent to the SDN controller, but all subsequent packets pass through the Video Detector and Policy Engines which directly control whether packets are sent out or sent to the Transcoder. Note that this allows SDNFV's Policy Engine to change the behavior of {\em all active} flows because every packet passes through it, not just the first two packets of each new flow like in the SDN case.

Our first video experiment compares the performance overhead for handling new flows between these two approaches. We use Pktgen-DPDK to generate a workload that varies the number of new incoming flows per second; after a flow has been established (i.e., it has sent two packets), it is replaced with a new flow. Figure~\ref{fig:expt:newflows} shows the output flow rates achieved by the SDN controlled system and the SDNFV controlled system. From the figure, we can see that the Controller quickly becomes the bottleneck when the input rate exceeds 1000 new flows/sec. On the other hand, the output rate of SDNFV can linearly increase, and achieves a maximum output rate 9 times greater, at which the SDN controller has become the bottleneck. The results show it is far more efficient to make dynamic flow decisions inside local NFs than frequently coordinating with the SDN controller.

Next we consider the flexibility with which these two approaches can adapt to changing conditions.  We configure the Policy engine NF and SDN Controller module so that for the period from 60 seconds through 240 seconds all packets will be sent to the video transcoder, which drops the traffic rate for each flow by half.  The workload mimics the behavior of 400 video flows, which each last for an average of 40 seconds before being replaced by a new flow.

\begin{figure}[t]
\begin{center}
\includegraphics[width=2.7in, trim=0.01cm 0.01cm 0.01cm 0.01cm, clip=true]{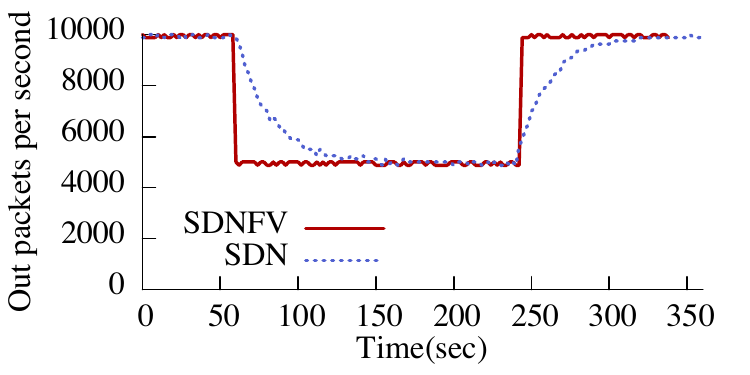}
\caption{SDNFV reacts quickly to policy changes, unlike an SDN that can only influence new flows.}
\label{fig:expt:dynamic}
\end{center}
\end{figure}

Figure~\ref{fig:expt:dynamic} shows the output rate of the video system over time.  Before the throttling period begins, the SDNFV Policy Engine NF examines each flow and issues a {\em ChangeDefault} message so when that flow is next seen by the Video Detector it will immediately be sent out the NIC instead of to the policy engine.  However, at t=60 seconds, the policy change causes the Policy NF to issue a {\em RequestMe} message, which causes all existing flows to be temporarily redirected to the policy engine, which can in turn adjust their default action so packets are sent to the Throttler NF.
In contrast, since the SDN controller based system only sends the first packets of a new flow to the SDN controller module, it can only add the transcoder NF to the path of new flows. This causes the SDN controller to significantly lag behind the target output rate, an effect that would be even more pronounced if flows had an average lifetime more than 40 seconds.

\subsection{Application Aware Flow Management}
\label{sec:eval:appaware}


There is a growing desire to perform network management not only on packet headers, but also layer-7 application data.
Our memcached-proxy NF is an application-aware load balancer that illustrates SDNFV's potential in this area.

The proxy parses incoming UDP memcached requests to determine what key is being requested.  The key is then mapped to a specific server using a hashing function, and the packet's header is rewritten to direct it to that server. We compare against the memcached proxy used by Twitter, which has understandably higher overheads since it uses interrupt driven packet processing and requires multiple packet data copies between kernel and user space. TwemProxy also needs to negotiate traffic in both directions since it maintains separate socket connections with the client and server. In contrast, Our NF-based proxy eliminates all packet copy overheads and can be seamlessly added to the network so it only intercepts incoming requests---the responses from the memcached server can be sent directly back to the client without any processing\footnote{TwemProxy also uses TCP instead of UDP like our proxy, however, we believe this does not significantly affect throughput or latency since our workload sends multiple requests over each TCP connection. Our proxy could support TCP flows by adding TCP hand-off capabilities~\cite{Mekky:2014:ADP:2620728.2620735}.}.

\begin{figure}[t]
	\begin{center}
		\includegraphics[width=2.7in, trim=0.01cm 0.01cm 0.01cm 0.01cm, clip=true]{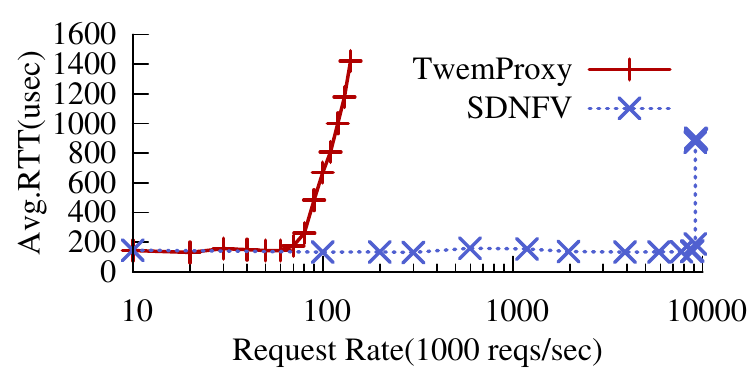}
		\caption{memcached RTT versus Request Rate}
		\label{fig:expt:requestRate-RTT}
	\end{center}
\end{figure}

To show the benefits of SDNFV, we measure the request round trip time as the get request rate is increased. Figure~\ref{fig:expt:requestRate-RTT} shows that TwemProxy quickly becomes overloaded when the rate is increased to only 90,000 req/sec. On the other hand, SDNFV can support 9,200,000 req/sec even with just one core, which is 102 times faster than TwemProxy. This illustrates the ``game-changing'' potential of NF-based services: they permit orders of magnitude higher performance and can be deployed in a flexible and transparent way.

\section{Related Work}

Among many challenges noted in the ETSI NFV standards~\cite{ETSI:NFVOverview:2014}, four research aspects: architectural design, NF performance, NF management, and NF placement, have widely been focused on.

{\bf Architecture:} Several projects are considering how SDN and NFV architectures must evolve. ONF  ~\cite{brief2014openflow} illustrates the idea that SDN can work with NFV and reveals the benefit. However, few demonstrated solutions exist as yet. Palkar et al.~\cite{Palkar:2015:EFN:2815400.2815423} rely on SDN centralized controller to manage the placement, service interconnection and dynamic scaling of a diverse of NFs. However, NFs and data plane lack the capability of dynamically steering the traffic based on traffic features.  All of the dynamic traffic steering across the NFs have to request SDN controller, which is expensive. To reduce the overhead of the controller, Ballani et al.~\cite{ballani2015enabling} propose data plane which has the capability of dynamically managing the traffic which has some similarity to us. This work only emphasizes the flexible flow management at the end hosts. We focus on the flexible traffic steering across the network. Packets still need to go through kernel, few information demonstrates how to get the high performance for NFs.
We propose a smart data and control plane architecture to efficiently manage different types of packet flows, which also provide high performance by zero-copy between NF and NIC \& across NFs located on the same server. Mekky et al.~\cite{Mekky:2014:ADP:2620728.2620735} propose a similar approach, but rely on data plane functions that are built into the Open vSwitch platform. We believe that running NFs in virtual machines provides better isolation and management; our approach also achieves significantly higher performance due to its zero-copy packet transfers. Cerrato et al.~\cite{Cerrato:FineGrainedNF:2014} support a fine-grained network functions through Intel DPDK. Sekar et al.~\cite{Sekar:2012:DIC:2228298.2228331} propose CoMB an architecture for efficiently consolidating middleboxes. SDNFV focuses on cross-layer flow management in the data and control planes.

{\bf NF Performance:} Packetshader~\cite{Han:2010:PGS:1851182.1851207}, netmap~\cite{rizzo_netmap:_2012}, NetVM~\cite{Hwang:2014:NHP:2616448.2616490,opennetvm2016}, and ClickOS~\cite{CLICKOS:NSDI:2014} all try to increase packet processing performance on COTS servers.
Netmap exposes a zero-copy packet API to user space applications to reduce packet processing overheads in non-VM environments. Packetshader uses a graphics processing unit (GPU) to off-load packets.  NetVM uses shared memory for efficient data transfer in inter-VM communication, whereas ClickOS modifies the virtualization layer to better support large numbers of independent, lightweight network function VMs. We built SDNFV on top of NetVM in order to minimize packet movement overheads, and we further optimize the platform to support parallel processing and service chain management.



{\bf NF Management:} Several NF control plane have been proposed to steer traffic between middleboxes and manage state such as PLayer~\cite{Joseph:2008:PSL:1402958.1402966}, SIMPLE~\cite{qazi2013simple}, Stra-tos~\cite{DBLP:journals/corr/abs-1305-0209} and Slick~\cite{Anwer:2013:SCP:2491185.2491223}.
Split/merge~\cite{Rajagopalan:2013:SSS:2482626.2482649} manages state for replicated virtual middleboxes to elastically scale based on workloads.  proposes system support for elastic execution in virtual middleboxes, and
Flowtags~\cite{Fayazbakhsh:2013:FEN:2491185.2491203} suggests network-wide policies for dynamic middlebox actions.
Similar to previous work, SDNFV is also motivated by adding more functions in the data plane and enriching existing SDN architecture. The key contribution of SDNFV is focusing on how to hierarchically manage flows across SDN controller, NF manager and individual NFs to achieve efficiency and flexibility.

{\bf NF Placement:}
VM placement has been studied for cloud computing~\cite{alicherry2013optimizing,applegate2010optimal}, but past work does not consider network characteristics which are key requirements for NFV service chains. Stratos~\cite{DBLP:journals/corr/abs-1305-0209} steers flows between available middleboxes by using an ILP formulation which considers a binary cost function between middleboxes. Placement of middleboxes is addressed by a rack-aware heuristic separately.
Traffic Steering systems like \cite{zhang2013steering} focus on inline service chaining and provide a simple heuristic for service placement, while \cite{qazi2013simple} provides an ILP formulation for traffic steering, but does not cover dynamic instantiation of middleboxes. Finally CoMb~\cite{Sekar:2012:DIC:2228298.2228331} places services along the pre-defined paths, so it lacks the dynamic nature of
placing NFVs in the network and the corresponding routing.

\section{Conclusion}
\label{section:conclusion}

We have described SDNFV, a flexible and dynamic architecture for high-speed software-based networking. To overcome the limitations of an overly simplified stateless data plane envisaged in current SDNs, 
SDNFV seeks to fully exploit virtualized network functions in enhancing 
the network's capabilities by incorporating a
hierarchical control framework to manage packet flows more flexibily. SDNFV still achieves both a high performance data plane and an easily managed control plane.

Our evaluation shows SDNFV can exploit complex state in the data plane without the overhead of frequently contacting the SDN controller. This provides far greater agility and precision compared to traditional SDN approaches that must proactively set coarse grained rules or only perform decisions based on the first packet in each flow.

\bibliographystyle{unsrt}
{\small
\bibliography{global-Eurosys}
}


\end{document}